# Generating contour lines using different elevation data file formats

P.S. Hiremath

Professor, Dept. of Computer Science, Gulbarga University, Gulbarga (KA) INDIA.

hiremathps@hotmail.com

B.G. Kodge

Lecturer, Dept. of Computer Science, S.V. College, Udgir, Dist. Latur (MH), INDIA

kodgebg@hotmail.com

## ABSTRACT

In terrain mapping, there are so many ways to measure and estimate the terrain measurements like contouring, vertical profiling, hill shading, hypsometric tinting, perspective view, etc. Here in this paper we are using the contouring techniques to generate the contours for the different digital elevation data like DEM, HGT, IMG etc. The elevation data is captured in dem, hgt and img formats of the same projected area and the contour is generated using the existing techniques and applications. The exact differences, errors of elevation (contour) intervals, slopes and heights are analyzed and recovered.

## Categories and Subject Descriptors

I.4.6 [Image Processing and Computer Vision]: Segmentation ¡V Edge and feature detection, Pixel classification

## General Terms

Algorithms.

## Keywords

Image processing, contouring with DEMs, 3d vector data, formats of elevation data.

## 1 INTRODUCTION

Contouring is the most common method for terrain mapping. Contour line connects points of equal elevation, the contour interval represents the vertical distance between contour lines, and the base contour is the contour from which contouring starts. Contour lines are lines drawn on a map connecting points of equal elevation. The contour line represented by the shoreline separates areas that have elevations above sea level from those that have elevations below sea level. We refer to contour lines in terms of their elevation above or below sea level. In this example, the shoreline would be the zero contour line ( it could be 0 ft., 0 m, or something else depending on the units we were using for elevation). Contour lines are useful because they allow us to show the shape of the land surface (topography) on a map. Suppose a DEM has elevation readings from 362 to 750 meters. If the base contour is set to 400 and the contour interval at 100, then contouring would create the contour lines of 400, 500, 600 and so on. Contour lines can be drawn for any elevation, but to simplify things only lines for certain elevations are drawn on a topographic map[1]. These elevations are chosen to be evenly spaced vertically. This vertical spacing is referred to as the contour interval. For example if the maps use a 10 ft contour interval, each contour lines are a multiple of 10 ft.( i.e. 0, 10, 20, 30, etc). Other common intervals seen on topographic maps are 20 ft (0, 20, 40, 60, etc), 40 ft (0, 40, 80, 120, etc), 80 ft (0, 80, 160, 220, etc), and 100ft (0, 100, 200, 300, etc). The contour interval chosen for a map depends on the topography in the mapped area. In areas with high relief, the contour interval is usually larger to prevent the map from having too many contour lines, which would make the map difficult to read.

The contour interval is constant for each map. It will be noted on the margin of the map. One can also determine the contour interval by looking at how many contour lines are between labeled contours.

Unlike the simple topographic map, the real topographic maps have many contour lines. It is not possible to label the elevation of each contour line. To make the map easier to read, every fifth contour line vertically is an index contour. Index contours are shown by darker brown lines on the map. These are the contour lines that are usually labeled [5].

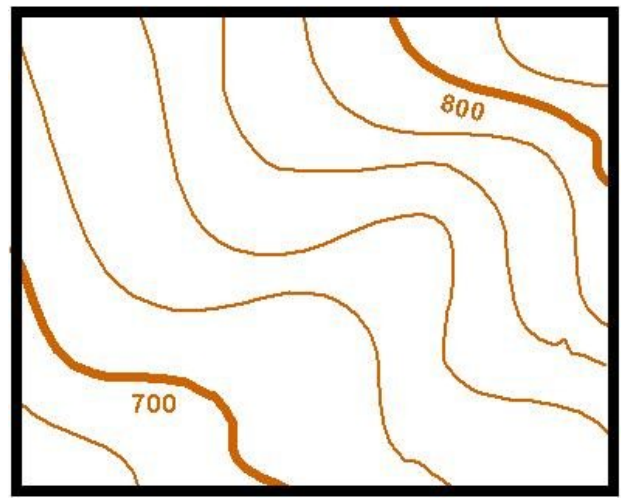

**Fig. 1 Section of a topographic map.**

The example Fig.1 illustrates a section of a topographic map. The brown lines are the contour lines. The thin lines are the normal contours, while the thick brown lines are the index contours. The elevations are only marked on the thick lines.

Because we only have a piece of the topographic map, we can not look at the margin to find the contour interval. But since we know the elevation of the two index contours, we can calculate the interval ourselves. The difference in elevation between the two index contours (800 - 700) is 100. We cross five lines as we go from the 700 line to the 800 line (note we don't include the line we start on but we do include the line we finish on). Therefore, we divide the elevation difference (100) by the number of lines (5). We will get the contour interval. In this case, it is 20. We can check ourselves by counting up by 20 for each contour from the 700 line. We should reach 800 when we cross the 800 line.

One piece of important information we can not determine from the contour lines on this map is the units of elevation. Is the elevation in feet, meters, or something else. There is a big difference between an elevation change of 100 ft. and 100 m ( 328 ft). The units of the contour lines can be found in the margin of the map. Most topographic maps in India use feet for elevation, but it is important to check because some do use meters.

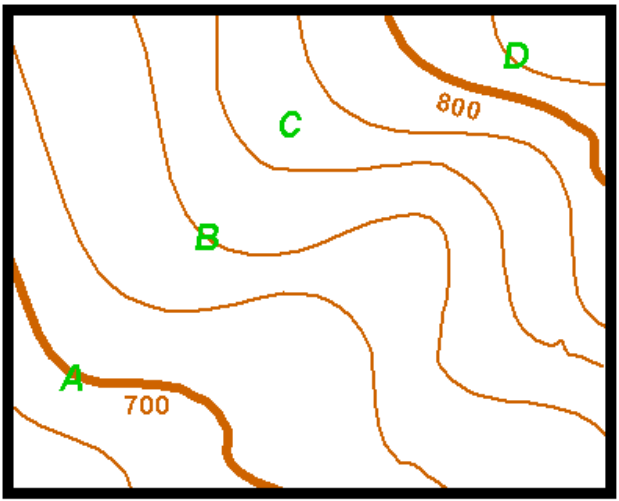

**Fig. 2 Estimated elevation of contour lines.**

Once we know how to determine the elevation of the unmarked contour lines, we should be able determine or at least estimate the elevation of any point on the map.

Using the Fig.2 we estimate the elevation of the points marked with letters A,B,C and D.

Point A = 700

Just follow along the index contour from point A until you find a marked elevation. On real maps this may not be this easy. One may have to follow the index contour a long distance to find a label.

Point B = 740

This contour line is not labeled. But we can see that it is between the 700 and 800 contour line. From above we know the contour interval is 20, so if we count up two contour lines i.e. (40) from 700, we reach 740.

Point C ~ 770

Point C is not directly on a contour line. But by counting up from 700 we can see that it lies between the 760 and 780 contour lines.

Because it is in the middle of the two, we can estimate its elevation as 770.

Point D = 820

Point D is outside the interval between the two measured contours. While it may seem obvious that it is 20 above the 800 contour, how do we know the slope hasn't changed and the elevation has started to back down? We can tell because if the slope stated back down we would need to repeat the 800 contour. Because the contour under point D is not an index contour, it can not be the 800 contour, so it must be 820.

## 2    READING HGT ELEVATION DATA

The elevation data in hgt, img and dem formats are captured from USGS(United States Geological Survey) and SRTM (Shuttle Radar Topographic Mission) sites for the region of Latur district in Maharashtra (INDIA) using Universal Transverse Mercator projection 18N076E Long/Lat with minimum 362 to maximum 700 meters of height is shown in Fig. 3[10]. These images are used for experimentation and validation of elevation data models in this paper.

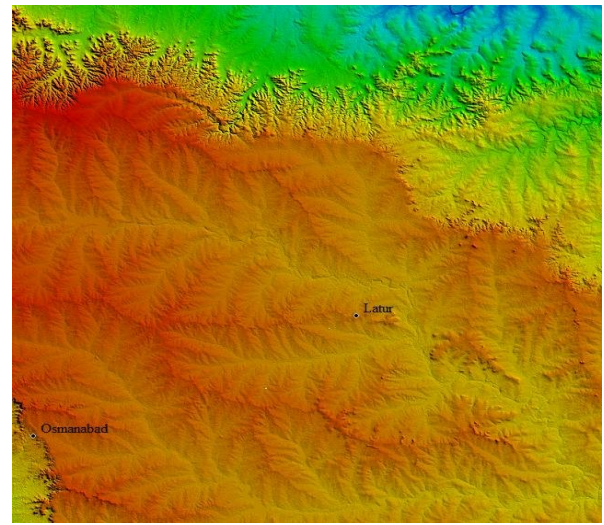

*Fig. 3 Elevation model of Latur region. (Courtesy USGS)*

The arrangement and pattern of contour lines reflect the topography. For example, contour lines are closely spaced in steep terrain and are curved in the upstream direction along a stream (Fig. 4). With some training and experience in reading contour lines, we can visualize and even judge the accuracy of the terrain as simulated by the digital data.

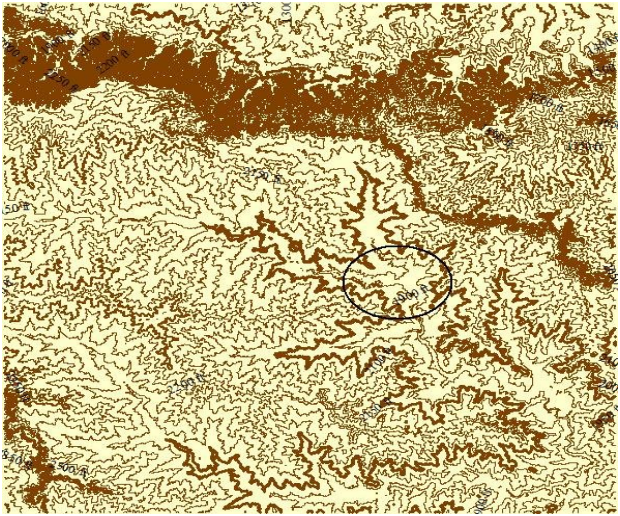

*Fig. 4. A contour line map with 50mtr interval using .hgt.*

Fig. 4 is an automatic contour map which is generated using Global mapper 10.0 with the elevation reading 362 to 750 meters and base contour generation is set to 400 with 50 meter contour interval.

Automated contouring follows two basic steps, namely, (1) detecting a contour line that intersects a raster cell or a triangle, (2) drawing the contour line through the raster cell or triangle.

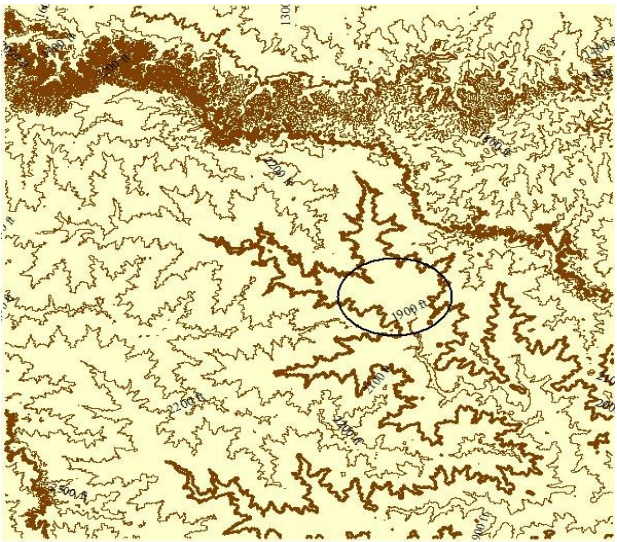

*Fig. 5 A contour map with 100 mtr interval using .hgt.*

The difference between Fig. 4 and Fig. 5 is highlighted by showing a circle at 1900 feet elevation. In Fig. 4 we can see the lower level contour (interval 50) lines showing the depth of surface area, but in Fig. 5 that particular contour lines are not generated due to the highest range of contour interval (ie 100 meters).

## 3    READING IMG ELEVATION DATA

In section II, we have already seen the contours of .hgt (Digital Elevation Model) elevation data file formats. Now we describe to image file formats (.img) of elevation data. The .img files with high bandwidth elevation data are some what same like .dem and .hgt file formats, but the actual resolution and clarity is poor than the .dem and .hgt due to the low band width. To differentiate the contour maps by using .hgt formats file and .img satellite imageries, consider map shown in Fig. 6.

The Fig. 6 uses the same parameters used as in Fig. 4 to generate the contour line map of the same projected area. But there are so many differences and errors that we may find out if we compare the Fig. 4 with Fig. 6.

The circles drawn in Fig. 6 are highlighting the major differentiation between the contour lines of the part of the same area in Fig. 4. The estimated height of both the contour lines of that area is same, but the contour lines are drawn differently .

The circles in Fig. 6 are errors which have occurred due to the insufficient data values available in the .img data formats.

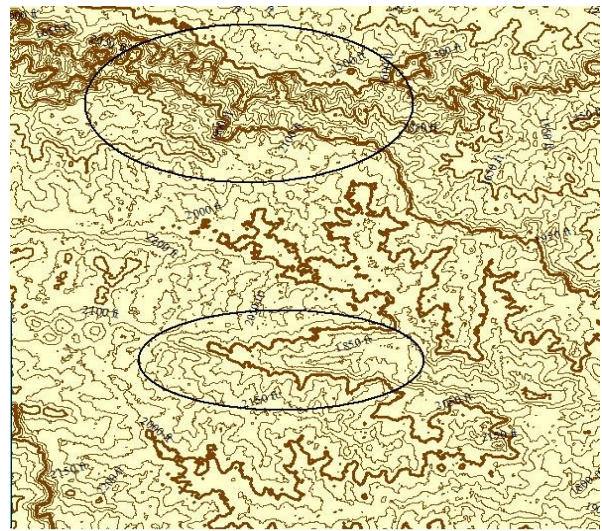

*Fig. 6 A contour line map with 50 mtr interval using .img*

## 4    ERROR DETECTION AND FIXING

After contour generation, many undesired line features are included in the overall contour vectors. Typical irregularities include contours without connection on either side, contours representing pure connection elements which do not contribute to a contour line continuation, or contour intersections. These irregularities are detected and eliminated step by step. Error checks have to be run several times, since the correction of one kind of error may induce another kind. Handling of these undesired contour features is focused on identifying and deleting arcs. Arc and node characteristics play the elemental role for the identification, which is based on the Info-Tables of elevation data and coordinate information.

For this purpose, typical topological characteristics for all node and arc classes are defined (Fig. 7):

- Dangle-Arc (DNG) – arc, which is at least on one end not connected to another arc.

- DeadEnd-Arc (DE) – dangle arc, where the connectivity is not given inside the data set boundaries.

- ArcShares2Polygons-Arc (AS2P) – arc, which contributes to two polygons representing the same elevation.

- PolygonInContour-Arc (PIC) – longer arc of two arcs forming a polygon, which is at the same time contributing to an unclosed contour line and therefore can cause confusion when it is interpreted.

- PolygonTouchPolygon-Arc (PTP) – longer arc of two arcs forming a polygon, which touches another polygon.

- Connection-Arc (CoA) – arc, which connects two or one polygon with an unclosed contour line and does not contribute to contour line continuation.

- Dangle-Node (DNG) – node, where an arc is not connected to another arc.

- DeadEnd-Node (DE) – dangle node, where an arc is not connected to another arc inside the data set boundaries.

- MultiArcConnection-Node (MAC) – node, where more than two arcs are connected.

- ClosedArc-Node (ClA) – node, represented by a closed arc (can be a MAC for one arc, while it represents a ClA for another closed arc at the same time).

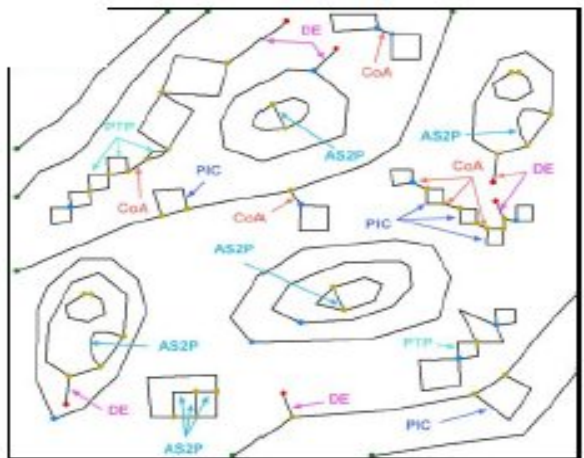

**6 Arc Classes:**
DNG: Dangle Arc
DE: Dead End Arc
AS2P: ArcShares2Polygons Arc
PIC: PolygonInContour Arc
PTP: PolygonTouch Polygon Arc
CoA: Connection Arc

**4 Node Classes:**
DNG: Dangle Node
DE: Dead End Node
MAC: MultiArcConnection Node
ClA: ClosedArc Node

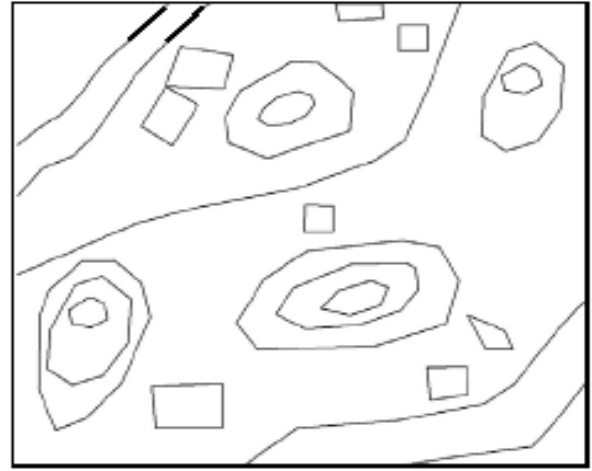

**Fig. 7(a) Error detection and fixing: Line scenario with highlighted arc and node classes (Top); (b) Contours after automated solving of undesired features (bottom).**

These defined arc and node classes are characterized by mutual relationship (topology). They can be identified by expanding, linking and analyzing entries from the associated attribute tables and lists, which are attached to the geometric elements in ArcInfo. Four tables are given: Node Attribute Table (NAT), Arc Attribute Table (AAT), Polygon Attribute Table (PAT), and Polygon Arc List (PAL). As an example, the node and arc classes 'Dead End' are described as follows. The topology characteristic of node class Dead End is defined by the 'sum of all incoming and outgoing arcs = 1'. There is either no incoming or outgoing arc associated with the node. For identification in ArcInfo the items 'Arc_per_Node' and 'Status' are added to NAT. The incoming and outgoing arcs in NAT are summed up and the results are written to the item 'Arc_per_node'. Then all nodes with 'final sum = 1' as well as the node coordinates (x, y) inside the data set boundaries are selected. Nodes in the NAT item 'Status' with 'DE' are attributed. The topology of the arc class 'Dead End' is characterized as arcs with a DE-node on one end and no connectivity to another arc. The location of this DE node is inside the data set boundaries. The detection of these arcs is performed by linking the node-IDs from the NAT to the 'FNODE#'-entry and 'TNODE#'-entry of the AAT. Then all arcs with the entry 'DE' in the NAT item 'Status' for at least one node are selected and deleted. After detection and deletion of a class, the nodes left along contours where there was a connection with a deleted arc are transformed to vertices during a topology update by joining single arcs, connected on either end to one other arc only[5].

## 5 APCONTOURS

The appearance and the quality of the contour lines depend strongly on the input data. Therefore, a high quality bare earth DTM without artifacts and noise processed with an appropriate grid size is required. The DTM resolution determines the level of surface detail and affects the appearance of the output contours. For example, a large cell size may result in coarse, blocky looking contour lines. Noisy DEMs produce contours of poor quality with a high level of details not representing the characteristics of the terrain. Intermap's core product Digital Terrain Model (DTM) produced with its airborne IFSAR technology ensures a high quality input dataset [4]. The grid cell size of 5 m as well as high level of horizontal and vertical accuracy allows the selection of very small contour intervals up to 2 feet (depending on the terrain and ground cover) as required for flood maps in flat areas. The seamless overlap of one pixel between adjacent DTM tiles guarantees contour continuation, since the APC process starts in the middle of the marginal pixels.

Contour lines produced by running APC process using Intermap's DTMs are called APContours. The elevation is stored as an attribute for each contour line according to the selected processing interval. The contours do not support breaklines or spot heights. They are free of common artifacts such as isolated or blocky lines (Fig. 8).

Intersections or Dead Ends exist if the interval is not appropriately selected to the terrain characteristics or manmade objects are contained in the DTM [1]. APContours of different intervals produced independently are identical at the same levels, e.g. contours of 5 m interval are identical with contours of 10 m interval at the same elevation.

APContours have been thoroughly tested using different intervals in feet and meters on more than 50 DTM tiles of NEXTMap USA in Geographic Coordinates and NEXTMap UK in OSGB projection. The tiles cover areas of different topography from flat to very steep terrain and different surface coverage. The tests showed that the APContours characterize the terrain very well in either case. Examples of the statistical results of the comparison of a certain number of arcs and nodes with the height of the input DTM for some NEXTMap USA data are listed. A high agreement between the APContours and the DTM can be noticed available either in Geographic Coordinate System or in UTM projection; the grid size was resampled to 5 m, except for OSGB DEM (10 m). Different contour intervals (between 1 and 10 m) were tested.

The quality of the input DEM largely determines the accuracy of the resulting contours. The comparison of APContours with contours generated from USGS DEM and high sample rate LIDAR data using the same processing parameters shows that APContours and USGS contours look quite similar unlike the LIDAR contours (Fig. 9). The LIDAR DEM exhibits much more noise and consequently the contours are rougher and include more errors like intersections.

Beside the statistical accuracy, the appearance of the contour lines is very important. To verify this, contours generated from Intermap's DTMs using two different commercial software packages (below called software package 1 and 2) as well as contour products distributed by other companies were compared to APContours. The overall appearance of contours processed using commercial software packages is quite similar compared to APContours, but the contours include some errors. They exhibit isolated small lines and many contour lines with spikes. In addition, the contours of software package 2 show Dead Ends, double lines and the run of the lines is quite blocky.

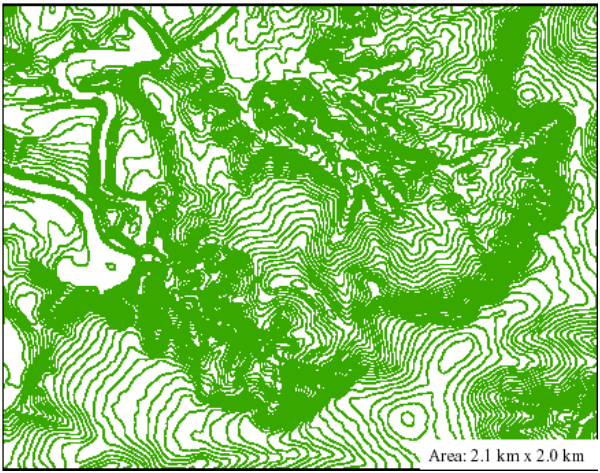

**Fig. 8: APContours; Interval: 2 ft**

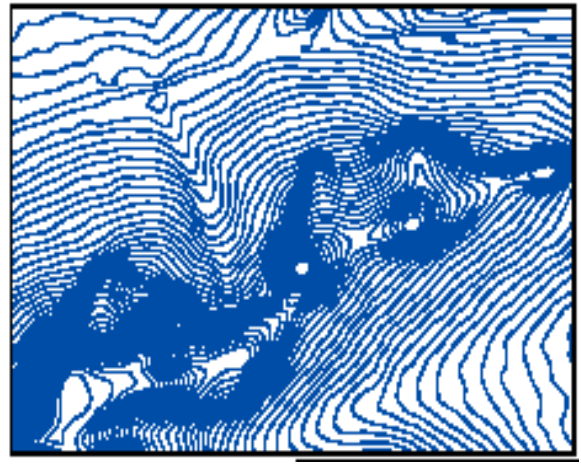

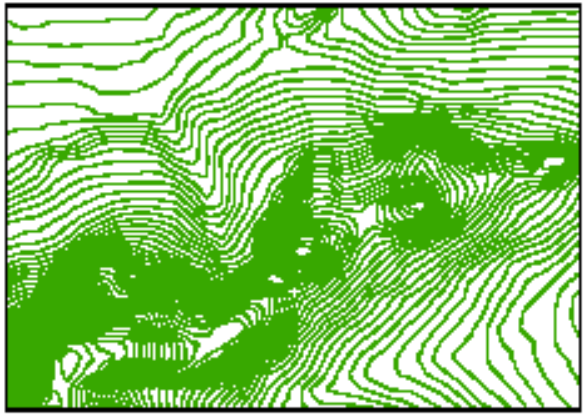

## 6 VALIDATION

For appraising the developed process and the resulting contour products, a series of tests were designed and conducted to validate both the APContours process as well as the Intermap's DTM as input. For this purpose the APC process was run using other DEMs as inputs and commercial contour programs were used to generate contour lines from Intermap's DTM. Furthermore, APContours were compared to contours distributed by other vendors of geo-data. For testing the robustness of the process, the APC process was used to generate contours from different input data like USGS (U.S. Geological Survey) DEM, LIDAR DEM generated from LIDAR point data, OSGB DEM (Ordnance Survey of Great Britain), and SRTM DEM. The DEMs are

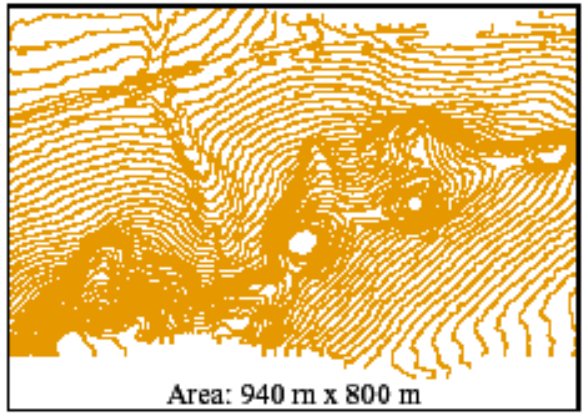

Fig. 9 Comparison: (a) APContours (blue), (b) contours generated from USGS DEM (green), and (c) from LIDAR DEM (orange); Interval: 1 m

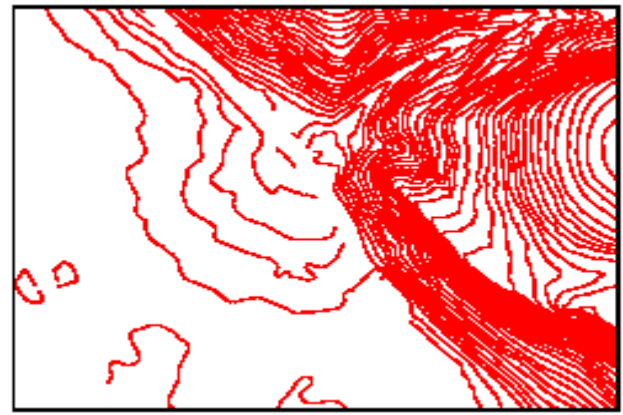

Fig. 10: Contour line continuity: APContours (blue) and vendor 1 contours (red); Interval: 5 m

As already mentioned, contour line continuation between separate generated tiles is very important. In Fig. 6 the results of the different software solutions are depicted. APContours show seamless continuation on tile edges.

Software package 1 generates interrupted contours and the contours generated using software package 2 show non fitting contour overlaps. For comparison of APContours with other commercial contour data products contours distributed by two different vendors of geo-data (following called vendor 1 and 2) were used. In the following example, the contours distributed by vendor 1 (interval: 5 m) cover a very flat coastal area. In the underlying OSGB DEM (grid size: 10 mtr) many man-made structures can be identified.

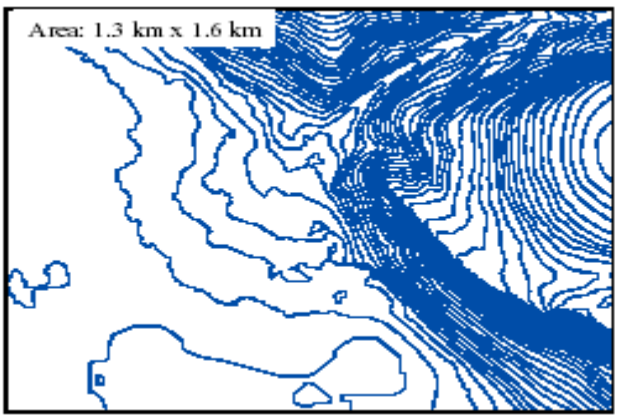

In general, the contours look smooth and appealing, but the continuity is not guaranteed. At some places, the lines are interrupted, whereas the APContours show continuous lines (Fig. 10). In some areas APContours appear blockier than the vendor 1 contours. Vendor 1 seams to generalize the contours stronger. That explains the worse result of the vendor 1 contours in the accuracy check.

## 7    CONCLUSION

The optimal selection of digital elevation data formats for better calculations plays very important role in the analysis and contour generation of a particular projected area. Because the pure output readings always depend upon the captured input data only. So a portable solution for automatically produced contour lines developed within ArcInfo environment and some other GIS packages are described. Applying its high quality DTMs and selecting an appropriate contour interval, Intermap has the capability to provide high quality contour lines on demand. The APContours are cartographically and aesthetically appealing and represent the shape of terrain very accurately. The high quality of the APContour product has been proved by comparing this data to contours generated using other commercial software and input DTMs as well as to other public-accepted contours.

## 8    REFERENCES.

## Author's Biography:

**1) Dr. P.S. Hiremath.**

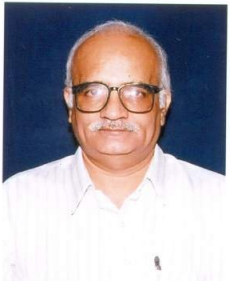

Dr. P.S. Hiremath is a Professor and Chairman, Department of P. G. Studies and Research in Computer Science, Gulbarga University, Gulbarga-585106 INDIA, He has obtained M.Sc. degree in 1973 and Ph.D. degree in 1978 in Applied Mathematics from Karnataka University, Dharwad. He had been in the Faculty of Mathematics and Computer Science of Various Institutions in India, namely, National Institute of Technology, Surathkal (1977-79), Coimbatore Institute of Technology, Coimbatore(1979-80), National Institute of Technology, Tiruchirapalli (1980-86), Karnatak University, Dharwad (1986-1993) and has been presently working as Professor of Computer Science in Gulbarga University, Gulbarga (1993 onwards). His research areas of interest are Computational Fluid Dynamics, Optimization Techniques, Image Processing and Pattern Recognition. He has published 102 research papers in peer reviewed International Journals and proceedings of conferences. Tel (off): +91 8472 263293, Fax: +91 8472 245927.

**2) Mr. Kodge B. G.**

Mr. Kodge Bheemashankar G. Working as a lecturer in department of studies and research center in Computer Science of Swami Vivekanand College, Udgir Dist. Latur

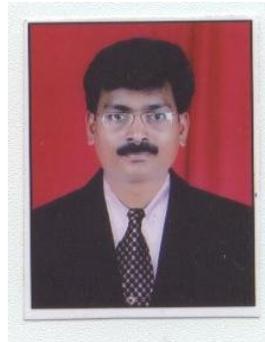

(MH) INDIA. He obtained MCM (Master in Computer Management) in 2004, M. Phil. in Comp. Sci. in 2007 and registered for Ph.D. in computer science in 2008. His research areas of interests are GIS and Remote Sensing, Digital Image processing, Data mining and data warehousing. He is published 10 research papers in national, international Journals and proceedings conferences. Tel. 09923229672.